\documentclass[aps,preprint,prd,showpacs,nofootinbib]{revtex4}
\usepackage{amsmath}
\usepackage{graphicx}
\usepackage{dcolumn}
\usepackage{bm}
\usepackage{amssymb}
\usepackage{latexsym}

\def\be{\begin{equation}}
\def\ee{\end{equation}}
\def\ba{\begin{eqnarray}}
\def\ea{\end{eqnarray}}

\begin{document}

\title{Conformally Dual to Inflation }

\author{Yun-Song Piao}

\affiliation{College of Physical Sciences, Graduate University of
Chinese Academy of Sciences, Beijing 100049, China}

\begin{abstract}

It is showed by a conformal rescaling that the inflationary background can be dual to a slowly expanding background,
which is almost Minkowski and described
by a conformal field theory conformally coupled to gravity.
It is proved that the primordial curvature perturbation and tensor perturbation generated during
these two conformally equivalent backgrounds are completely equal,
and the scale invariance of perturbations
is determined by the conformal invariance of field theory in slowly expanding background.
In dual slowly expanding scenario, the primordial universe
is asymptotical to a static state in infinite past. We discuss the
implication of the results obtained.


\end{abstract}

\maketitle

The inflation scenario is the current paradigm of the early universe,
which is highly favored by observations. However, the inflatons are diversified in diversified
inflation models \cite{Mazumdar}. What is inflaton is still
a significant issue. Here, it will be disposed in an insightful perspective.

The perturbation on large scale consists of a constant mode and a mode
dependent of time \cite{Mukhanov}.
When the scale factor is rapidly changed while
$\epsilon\ll 1$ is nearly constant, the constant mode is responsible
for the scale invariance of the primordial perturbations in inflation scenario\cite{MC}.


However, the
scale invariance of curvature perturbation can also be obtained for $|\epsilon|\gg 1$.
The evolution with
$\epsilon\gg 1$ is the slowly contracting, which is that of
ekpyrotic universe \cite{KOS}. While $\epsilon \ll -1$ gives the
slow expansion \cite{PZhou}, which has been applied for island
universe \cite{island}.
In slowly evolving scenario,
the tensor perturbation
generated is generally strongly blue, which implies that it is negligible on large scale. While
in inflationary scenario, the tensor perturbation is scale invariant, which can
be detected.

Recently, it has been found in \cite{Piao1109} that, in a background of slowly expanding with the rapidly changed gravitational coupling, both the
curvature perturbation and the tensor perturbation can be scale
invariant, and the ratio of the tensor to scalar is in a
regime consistent with that of inflation scenario.
This result is actually a reflection
that the perturbation is
conformal invariant fully nonperturbatively \cite{Gong}.
The conformal invariance
of perturbations implies that the background evolution having
coincident predictions with that of inflation might be
designed by a conformal rescaling of inflationary background.
Thus it might be significant to contemplate inflation in its conformally
dual background, which might bring a fruitful insight to the
inflation scenario itself and also the early universe.

Here, we will show that the inflationary background can be conformally dual to a slowly expanding background, which is described
by a conformal field theory conformally coupled to gravity.
We prove that the primordial curvature perturbation and tensor perturbation
generated in both pictures are be completely equal, and
the scale invariance of perturbations
is determined by the conformal invariance of field theory in slowly expanding background.
We discuss the
implication of the results obtained.


We firstly will show the slowly expanding background, which is described
by a conformal field theory conformally coupled to gravity, is actually conformally dual to the inflationary background.
We begin with
\be {\cal L}\sim {M_{Peff}^2\over 2}R+{1\over
2}(\partial_\mu \varphi)^2-{1\over 4}\lambda \varphi^{4-\Delta} ,
\label{L}\ee where $M_P^2=1$, $M_{Peff}^2=\xi \varphi^2$, and $\xi\simeq 1/6$ will
be determined.
In \cite{Rubakov}, the perturbation of the field conformally coupled to the gravity
has been studied, however, there the field is a light field which
hardly affect the cosmological evolution, and the potential of field is negative.
Here, the potential is positive, and the field is ghostlike, which is
required for implementing the slow expansion. However,
as will be
showed, there is not the ghost instability.

The calculation is highly similar to that in \cite{PZhou}.
The background evolution of the slow expansion is
\cite{PZhou},\cite{Piao0404}, \be a\sim {1\over (t_{*}-t)^{{1/
|\epsilon|}}},\,\,\,\,\, H= {1\over |\epsilon|(t_{*}-t)}
\label{H13}\ee for $\epsilon\ll -1$.
When $3H^2\varphi^2 \ll 6H{\dot \varphi}\varphi$ is neglected, the Fridmann equation
is simplified as \be H\simeq {-{1\over 2}{\dot \varphi}^2 + {1\over 4}\lambda
\varphi^{4-\Delta}\over 3{d \over dt }M_{Peff}^2}\sim {1\over
|\epsilon|(t_{*}-t)}. \label{HH}\ee This requires $\Delta\equiv {2\over |\epsilon|}$, and \be \varphi\sim
\left(1+{\cal O}({1\over |\epsilon|})\right) {\sqrt{2/
\lambda}\over (t_*-t)^{1/|\epsilon|+1}}, \label{phi1}\ee
where the
deviation $\sim 1/|\epsilon|$ is required to accurately give
Eq.(\ref{HH}). Thus $\varphi\sim {1\over (t_*-t)}$ for
$\epsilon\ll -1$. The result is consistent with $3H^2\varphi^2 \ll
6H{\dot \varphi}\varphi$, since \be H\varphi \sim {1\over
|\epsilon|(t_*-t)^2}\ll {\dot \varphi} \sim {1\over (t_*-t)^2}
\label{ll}\ee for $\epsilon\ll -1$.
The equation of $\dot H$ is given by,
\be 2M_{Peff}^2{\dot H}-\left({d \over dt}M_{Peff}^2\right)H= {{\dot \varphi}^2
-{d^2 \over dt^2}M_{Peff}^2}, \ee where $M_{Peff}^2=\xi\varphi^2$.
This requires
\be \xi\simeq {1\over 6}(1+{1\over 3|\epsilon|}). \label{xi} \ee
Thus for $|\epsilon|\gg 1$, $\xi=1/6$ is just that of
conformal coupling. This implies that the field theory (\ref{L}) is a conformal field theory,
conformally coupled to the gravity, however, there is a slight deviation of $\sim 1/|\epsilon|$. We will see it is this
deviation that determines the tilt of the spectrum of primordial perturbation.

We conformally rescale $g_{\mu\nu}$ as \be g_{E\mu\nu}=M_{Peff}^2g_{\mu\nu}.\ee
We follow Ref.\cite{FT}. The line element is $ds^2_E=M_{Peff}^2 ds^2$, which gives $dt_E=M_{Peff} dt$. Thus we have
\be t_E=\int M_{Peff} dt={|\epsilon|\over \sqrt{3\lambda}(t_*-t)^{1/|\epsilon|}}, \label{tE}\ee where $t_E$ is positive and Eq.(\ref{xi})
is applied. $a_E=M_{Peff}a$ brings $H_E$,
\be H_E ={1\over M_{Peff}}\left(H+{{\dot M}_{Peff}\over M_{Peff}}\right)
\simeq
{|\epsilon|\over t_E}, \label{HE}\ee where Eq.(\ref{tE}) is applied. Thus $a_E$ is given by
$a_E\sim exp{\int H_E dt_E}$, which is \be a_E \sim t_E^{|\epsilon|}. \label{aE}\ee This is just inflation since
$|\epsilon|\gg 1$. The field $\varphi_E$ after this conformal rescaling is canonical \be
d\varphi_E^2
= {1\over M_{Peff}^2}\left({6 M_{Peff,\varphi}^2}-1\right)
d\varphi^2 \simeq
{2\over |\epsilon|\varphi^2} d\varphi^2. \ee Thus
$\varphi_E=\sqrt{2\over |\epsilon|}{\ln{\varphi}}$.
Thus (1) after the conformal rescaling is given by
\be {\cal L}_E\sim {1\over 2}R_E-{1\over 2}(\partial_{E\mu}\varphi_E)^2-9\,\lambda
\,e^{-\sqrt{2\over |\epsilon|}\varphi_E},\ee
which is actually consistent with Eq.(\ref{aE}). Thus it can be found
\be |\epsilon|={1/ \epsilon_E}. \label{epsilon}\ee
Thus the slowly expanding background $|\epsilon|\gg 1$, which here is
described by a conformal field theory conformally coupled to gravity, is conformally dual to the inflationary background $\epsilon_E\ll 1$.





The curvature perturbation $\cal R$ and the tensor perturbation $h_{ij}$ are
actually conformal invariant. Here, we will prove it in detail, which helps to
clarify what determines the scale invariance of the perturbations.

In inflationary background $a_E\sim t_E^{1/\epsilon_E}$, the results of the primordial perturbations generated are familiar,
which have been listed in Table I. In its conformal dual background given
by Eq.(\ref{H13}), the case is slightly complicated. We will calculate
$\cal R$ and $h_{ij}$ in this picture.
The quadratic action of curvature perturbation ${\cal R}$ for the most
general single field, including the nonminimal coupling case \cite{Qiu},\cite{Tsu1}, is \cite{KYY3}
\be S_2\sim \int d\eta d^3x {a^2 Q_{\cal R}\over c_{\cal
R}^2}\left({{\cal R}^\prime}^2-{c_{\cal R}^2}(\partial {\cal
R})^2\right), \label{LR}\ee where $Q_{\cal R}>0$
and $c_{\cal R}^2>0$ are required to avoid the ghost and gradient
instabilities.
Thus the equation of $\cal R$ is given by \cite{Muk},\cite{KS},
\be u_k^{\prime\prime} +\left(c^2_{\cal R}
k^2-{z^{\prime\prime}_{\cal R}\over z_{\cal R}}\right) u_k = 0, \label{uk}\ee  after
$u_k \equiv z_{\cal R}{\cal R}_k$ is defined, where $z_{\cal R}= \sqrt{2}{a {Q}_{\cal
R}^{1/2}/ c_{\cal R}}$ and the prime is the
derivative for $\eta$.

The scale invariance of $\cal R$ requires \ba z_{\cal R}\sim {a
Q_{\cal R}^{1/2}\over c_{\cal R}} &\sim & {1\over \eta_*-\eta}\,\,
{for}\,\, {{\rm constant}}\,\,{ {\rm mode}}
\label{z2}\\
&or & (\eta_*-\eta)^2 \,\, {for}\,\,{ {\rm increasing}}\,\,{{\rm
mode}} \label{z1}\ea has to be satisfied.
In certain sense, both evolutions are dual \cite{Wands99}.
Here, $c_{\cal R}^2$ is not changed. However, the changed $c_{\cal R}^2$ will
alter the result \cite{Picon},\cite{Piao0609},\cite{JM},\cite{KP},\cite{Kinney},\cite{Kinney1},\cite{JK1},\cite{K2}.
During slow evolution, the scale
invariant curvature perturbation can be induced
by either its constant mode \cite{KS1},\cite{LMV},\cite{KM},\cite{JK}, or its increasing mode \cite{Piao1012},\cite{Piao1105},
In inflation scenario, the scale invariant perturbation is induced by
the constant mode of perturbation, thus in its dual slow expansion, the perturbation
certainly must be that induced by the constant mode.



Here, $c_{\cal R}^2=1$ and $Q_{\cal R}= {1\over 6|\epsilon|}\varphi^2>0$
\cite{Piao1109}. Thus
$z_{\cal R}$ is \ba z_{\cal R}& = & \sqrt{2}{a {Q}_{\cal
R}^{1/2}\over c_{\cal R}} \sim {\sqrt{2} \over
\sqrt{3\lambda}|\epsilon|(t_*-t)^{2/|\epsilon|+1}} \nonumber\\ &\sim &{\sqrt{2} \over
\sqrt{3\lambda}|\epsilon|(\eta_*-\eta)^{1/|\epsilon|+1}}, \ea where
$t\sim \eta^{1-1/|\epsilon|}$ for $\epsilon\ll -1$.
When $k^2\gg z_{\cal R}^{\prime\prime}/z_{\cal R}$, the perturbation is deep
inside the $\cal R$ horizon $1/{\cal H}_{freeze}^{\cal R}$, $u_{{\cal R}}\sim {1\over \sqrt{2k}}
e^{ik\eta}$.
Here, the $\cal R$ horizon
\be {\cal H}^{{\cal R}}_{freeze}= \sqrt{z^{\prime\prime}_{{\cal R}}\over z_{{\cal R}}}
\sim {1\over \eta_*-\eta}={\cal H}^{{\cal R}}_{Efreeze}
\ee
which is actually a reflection
of the conformal invariance of $z_{{\cal R}}(\eta)$.
When $k^2\ll z^{\prime\prime}/z$,
\be u_{\cal R}\simeq {1\over \sqrt{2k}}{1\over (-k\eta)^{1/|\epsilon|+1}}
.\ee
Thus ${\cal P}_{\cal R}$ is given by \be {\cal P}_{\cal R}={k^3\over 2\pi^2} \left|{u_{{\cal R}}\over z_{\cal R}}\right|^2
\simeq {3\over 8\pi^2}\lambda|\epsilon|.\label{pr}\ee
The tilt
is $n_{\cal R}\simeq 1-{2\over |\epsilon|}$. Thus the only adjusted
parameter $\lambda$ of the model is normalized by ${\cal P}_{\cal R}\sim 1/10^{10}$, and there is not additional finetuning.


The quadratic action of the tensor perturbation $h_{ij}$ is \be
S_2\sim \int d\eta d^3x {a^2 Q_{T}\over
c_{T}^2}\left({h_{ij}^\prime}^2-{c_T^2}(\partial
{h_{ij}})^2\right), \label{LT}\ee 
where $Q_{T}>0$ and $c_{T}^2>0$ are required to avoid the ghost
and gradient instabilities. The shape of this action is similar to that
of the curvature perturbation.



\begingroup
\begin{table}
    \label{spectrum}
    \begin{tabular}{|c|c|c|c|c|}
      \hline
     \multicolumn{5}{|c|}{conformal dual backgrounds}
       \\ \hline \hline   & \multicolumn{2}{|c|}{inflation $\epsilon_E\ll 1$ } &
       \multicolumn{2}{|c|}{slow expansion $|\epsilon|\gg 1$ } \\
      \hline \hline
      $n_{\cal R}$  & \multicolumn{2}{|c|}{$ 1-2\epsilon_E$} &
       \multicolumn{2}{|c|}{$1-{2\over  |\epsilon|}$}
       \\
      \hline  \hline
      $r$  & \multicolumn{2}{|c|}{$16\epsilon_E$} &
       \multicolumn{2}{|c|}{${16\over |\epsilon|}$ }\\
      \hline  \hline
      $n_T$  & \multicolumn{2}{|c|}{$-2\epsilon_E$} &
       \multicolumn{2}{|c|}{$-{2\over |\epsilon|}$ }\\
      \hline \hline
    \end{tabular}
    \caption{The $n_{\cal R}$, $r$ and $n_T$ for
    inflation with $\epsilon_E\ll 1$ and slow expansion $|\epsilon|\gg 1$, respectively.
    Here, the conformal rescaling by which the slow expansion is dual to inflation
    implies $|\epsilon|=1/\epsilon_E$.
     }
  \end{table}
  \endgroup

The spectrum of $h_{ij}$ is determined by $Q_T$, which is
given by $Q_T=M_{Peff}^2\simeq {\varphi^2/ 6}$, and $c_T^2=1$ \cite{Piao1109}. Thus similarly \be z_T= 0.5\,{a
Q_{T}^{1/2}\over c_{T}}\sim {1\over
2\sqrt{3\lambda}(\eta_*-\eta)^{1/|\epsilon|+1}}. \ee
When $k^2\gg z_{T}^{\prime\prime}/z_{T}$, the perturbation is deep
inside the $h_{ij}$ horizon $1/{\cal H}_{freeze}^{T}$, $u_{T}\sim {1\over \sqrt{2k}}
e^{ik\eta}$.
Here, the $h_{ij}$ horizon
\be {\cal H}^{T}_{freeze}= \sqrt{z^{\prime\prime}_{T}\over z_{T}}
\sim {1\over \eta_*-\eta}={\cal H}^{T}_{Efreeze}
\ee
which is actually a reflection
of the conformal invariance of $z_{T}(\eta)$. When $k^2\ll z^{\prime\prime}/z$,
\be
u_{T}\simeq {1\over \sqrt{2k}}{1\over (-k\eta)^{1/|\epsilon|+1}}.\ee
Thus ${\cal P}_T$ is given by
\be
{\cal P}_T= {k^3\over \pi^2}\left|{u_{T}\over z_{T}}\right|^2\simeq {6\over \pi^2}\lambda. \ee The tilt is $n_T\simeq -{2\over |\epsilon|}$.
Thus the ratio of tensor to scalar $r\equiv {{\cal
P}_T\over {\cal P}_{\cal R}}$ is given by \be r\simeq {16\over |\epsilon|}, \label{r}\ee which is only dependent of
$|\epsilon|$. Here, if (\ref{epsilon}) is applied for Eq.(\ref{r}),
$r$ obtained is just that of inflation with
constant $\epsilon_E$.
Thus the inflation
and the slow expansion have equal predictions, which are listed in Table I.

Thus it is obviously found that the scale invariance of primordial perturbations
is determined by the conformal invariance of field theory in slowly expanding background. However,
this conformal field is actually a slightly deformed conformal field, there is a slight deviation of $\sim 1/|\epsilon|$ from
the conformal invariance, which naturally brings a slight red tilt of the primordial perturbation.

In \cite{Rubakov},\cite{HK}, the similar conformal field
has been applied, though for a picture of slow
contraction. However, in \cite{Rubakov},
the field conformally coupled to the gravity is a light field, which
hardly affect the cosmological evolution. While in \cite{HK},\cite{CNT}, though the conformal field
is dominated, it is not conformally coupled with the gravity. Thus both fail to
conformally dual to inflation. In \cite{Rubakov},\cite{HK},\cite{CNT}, the conformal invariance
is only related to
the scale invariance of the perturbation of a light field. While here what the conformal invariance
determines is that of metric perturbations, which is adiabatically generated.





\begin{figure}[t]
\begin{center}
\includegraphics[width=6cm]{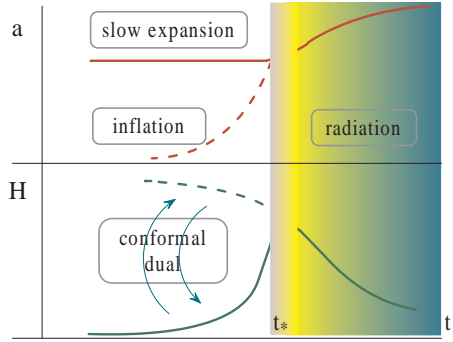}
\caption{ The evolutions of $a$ and $H$ in inflation picture (dashed line) and
its conformally dual, i.e.slow expansion picture (solid line), respectively.
Before $t_*$, which is $t_{E*}$ for inflation, due to the rapidly change of $M_{Peff}$,
both evolutions are completely different, in which $t$ is from -infinity to $t_*$ while
$t_E$ is from 0 to $t_{E*}$, in term of Eq.(9), noting that the conformal time
$\eta$ is same for both evolutions, which is from -infinity to $\eta_*$. However, after $t_*$,
since $M_{Peff}=M_P$ is fastened, both evolutions coincide.
$t_*$ is the heating time,
after which the evolution of hot big bang model begins.
 }
\end{center}
\end{figure}

Though the predictions of inflation picture and its
dual slow expansion picture for the perturbations are indistinguishable, the evolutions of their backgrounds
are completely different.
The inflation spacetime is geodesically incomplete,
which thus is initially singular \cite{BGV}.
However, in its conformally dual background, the case is different,
the primordial universe is
slowly expanding, which is asymptotical to a static state in infinite past,
and thus is singular free. This is illustrated in Fig.1.

Thus this
conformal duality might imply an alternative to the origin of
observational universe. The primordial universe might be in a slowly expanding conformal phase, in which
the full theory depicting the universe is conformal invariant.
As a result of the conformal invariance
of underlying theory, both the curvature perturbation and tensor perturbation
emerged in this conformal
phase are naturally scale invariant. When the conformal phase ends, the energy of conformal field will be
released to heat the universe. Here, the heating mechanism might be similar to
that after inflation.
However, after
the heating, $M^2_{Peff}= M_P^2$ should be fastened, or there will be conflicts with observations,
which will be studied elsewhere.
Hereafter, the evolution of universe coincides with that of hot big bang model, see Fig.1.

Here, the picture is as if homological with pre big bang picture \cite{GV},\cite{PBB}.
However,
there the superinflationary expanding phase in string frame, which but is not the slow
expansion, is conformally dual
to a contracting phase. While here the conformal
rescale of the slow expanding phase is inflation, thus the scale invariance of the curvature perturbation and
the tensor perturbation is simply obtained.


Here,
with Fig.1, it might be designed
that this infinite past could be jointed to the infinite latetime of
previous universe. In principle, after doing this joint an infinite number of times,
we will have a cyclic universe. Here, the infinite past is actually a conformally dual to
inflation initial singularity. In this sense, this conformal duality might imply
a feasible implementing of the conformal cyclic cosmology \cite{Penrose},
which is left
for the aftertime.



In conclusion, we show that the inflationary background can be conformally dual
to a slowly expanding background.
We prove that the primordial curvature perturbation and tensor perturbation
generated during
both backgrounds are completely equal, and
the scale invariance of perturbations
is determined by the conformal invariance of field theory in slowly expanding background.
Though we only consider a simple case, we conjecture that all inflation models
can be conformally dual to a slightly deformed conformal field theory in
slowly expanding background, which might imply that though the inflatons are diversified in diversified
inflation models, all actually have a common origin, when being performed in its conformally dual background.
In dual slow expansion,
the primordial universe is asymptotical to a static state in infinite past, and thus is singular free.
Here, this slowly expanding phase is
depicted by a conformal field theory, thus it might has a dual description of AdS$_5$ gravity \cite{CFT},
which might be interesting for studying.








\textbf{Acknowledgments} We thank Yi-Fu Cai, Taotao Qiu for discussions. This
work is supported in part by NSFC under Grant No:10775180,
11075205, in part by the Scientific Research Fund of
GUCAS(NO:055101BM03), in part by National Basic Research Program
of China, No:2010CB832804.

\end{document}